\documentclass[aps,prd,showpacs,twocolumn]{revtex4}

\begin{document}

\def\be{\begin{equation}}
\def\ee{\end{equation}}
\def\bd{\begin{displaymath}}
\def\ed{\end{displaymath}}
\def\ba{\begin{eqnarray}}
\def\ea{\end{eqnarray}}
\def\lr{\leftrightarrow}
\def\ccbar{$c\bar c$ }
\def\td{$1^3{\rm D}_J$ }
\def\tdth{$1^3{\rm D}_3$ }
\def\tdt{$1^3{\rm D}_2$ }
\def\tdo{$1^3{\rm D}_1$ }
\def\sdt{$1^1{\rm D}_2$ }
\def\S{\rm S}
\def\L{\rm L}
\def\J{\rm J}
\def\B{\rm B}
\def\D{\rm D}
\def\P{\rm P}
\def\K{\rm K}
\def\X{\rm X(3872) }

\title{Charmonium Options for the X(3872)} 
\author{Ted Barnes$^a$\footnote{Email: tbarnes@utk.edu}
and Stephen Godfrey$^b$\footnote{Email: godfrey@physics.carleton.ca}}
\affiliation{
$^a$Department of Physics, University of Tennessee,
Knoxville, TN 37996,
USA,\\
Physics Division, Oak Ridge National Laboratory,
Oak Ridge, TN 37831, USA}
\affiliation{
$^b$Ottawa-Carleton Institute for Physics, 
Department of Physics, Carleton University, Ottawa, Canada K1S 5B6 }

\date{\today}

\begin{abstract}
In this paper we consider all possible 1D and 2P \ccbar assignments for the 
recently discovered X(3872). Taking the experimental mass
as input, we give numerical results for the E1 radiative widths
as well as the three principal types of strong decays; open-charm, \ccbar
annihilation and closed-charm hadronic transitions. We find that many 
assignments may be immediately eliminated due to the small observed total 
width. The remaining viable \ccbar assignments are 
$1^3$D$_3$,
$1^3$D$_2$,
$1^1$D$_2$,
$2^3$P$_1$ and
$2^1$P$_1$. A search for the mode $J/\psi\, \pi^o \pi^o$ can establish the
C-parity of the X(3872), which will 
eliminate many of these possibilities.
Radiative transitions can then be used to test the remaining  
assignments, as they populate characteristic final states.
The $1^3$D$_2$ and $1^1$D$_2$ states 
are predicted to have large ({\it ca.}50\%) 
radiative branching fractions to $\chi_{c1}\gamma$ and $h_c\gamma$ 
respectively. We predict that the $1^3$D$_3$ will also be
relatively narrow and will have a significant ({\it ca.}10\%) 
branching fraction to $\chi_{c2}\gamma$, and should also 
be observable in B decay.
Tests for non-\ccbar \X assignments are also discussed.

\end{abstract}
\pacs{12.39.-x, 13.20.Gd, 13.25.Gv, 14.40.Gx}

\maketitle

\section{Introduction}

Several new mesons states have recently been reported 
\cite{Cho03,Bau03,Aub03,Bes03,Kro03}
whose properties are in disagreement with the predictions of 
quark potential models. 
Assuming experimental confirmation, this
indicates the necessity of refinements in the models
or the inclusion of additional dynamical effects.  

The most recent of these discoveries is the  
X(3872), which was reported by the Belle Collaboration \cite{Cho03}
in the $J/\psi\,\pi^+\pi^- $
invariant mass distribution in the process 
${\B}^\pm \to {\K}^\pm J/\psi\,\pi^+\pi^- $. The mass and width
upper limit reported by Belle are
\begin{equation}
{\rm M} = 3872.0\ \pm \; 
0.6\; {\it (stat)} \ \pm \; 0.5\; {\it (sys) }\  {\rm MeV},
\end{equation}
\begin{equation}
\Gamma^{tot.}_{\X}\; < \; 2.3\ {\rm MeV}\ \ 95\% \ {\rm C.L.} %{\it c.l.}
\end{equation}
Note that the mass is very near the 
${\D}^o {\D}^{*o}$ threshold of $3871.5 \pm 0.5$~MeV.
The width is 
consistent with experimental resolution.  
This observation has since
been confirmed by CDF \cite{Bau03}, who report
a very similar mass of
\begin{equation}
{\rm M} = 3871.4\ \pm \;
0.7\; {\it (stat)} \ \pm \; 0.4\; {\it (sys) }\  {\rm MeV}
\end{equation}
for a fixed experimental resolution of 4.3~MeV.
A limit on a relative radiative branching fraction
has also been reported \cite{Cho03},
\begin{equation}
{{\B}({\X}\to \chi_{c1} \gamma)\over 
{\B}({\X}\to J/\psi\,\pi^+\pi^- )} \ < \ 0.89, \ \ 90\% \ {\rm C.L.} 
\end{equation}
There was a prior, unconfirmed, observation of a $2^{--}$ state, 
$\psi(3836\pm 13)$ in $\pi^\pm N \to J/\psi \pi^+\pi^- + 
\hbox{anything}$ by Fermilab E-705 \cite{e705}.
  
An obvious assignment for the \X would be
an L=2 \ccbar level, since the \tdt and \sdt
states are both expected to be narrow due to the absence of 
DD decay modes \cite{Abbrev}
and are expected to have sizeable production rates in $B$-decays
\cite{Chi98,Yua97,Ko97}.
These assignments however have the problem that the
mass of the \X is somewhat higher than most potential models predict 
for 1D \ccbar states (see Table~I). Another difficulty is that
the Belle limit on the relative branching fraction
${\rm B}_{\chi_{c1}\gamma } / {\rm B}_{J/\psi\,\pi^+\pi^- }$   
is much smaller than the ratio predicted by
Eichten {\it et al.} \cite{Eic02}
for the C = (+) \tdt state,
although this may simply be due to an inaccurate estimate of the 
problematic rate to $J/\psi\,\pi^+\pi^- $. 

These difficulties have led to speculations that the 
\X may not be a conventional 1D \ccbar state.
The proximity to the DD$^*$ threshold in particular has suggested  
that the \X might be a weakly-bound DD$^*$ molecule 
\cite{Tor03,Clo03b,Vol03b,Pak03,Won03,Bra03,Swa03}.
Other possibilities that have been discussed are a
2P \ccbar state \cite{Pak03,Clo03b} or a charmonium hybrid 
\cite{Clo03a,Clo03b}.

In this note we compare the properties of the 
\X to theoretical predictions for the radiative transitions
and strong decay rates of all 1D and 2P charmonium states.
We begin by summarizing quark model 
predictions for the masses of the 1D and 2P 
\ccbar states,   
followed by our predictions for radiative 
transitions and strong decay partial widths.  
From these results we determine which
\ccbar assignments appear consistent with the experimental data at present,
following which we suggest measurements that can differentiate between 
these \ccbar assignments as well as non-$c\bar c$ possibilities.  

\eject

\section{Spectroscopy}

The spectrum of charmonium states has long provided important tests
of our understanding of the forces between quarks. The mean multiplet
positions are consistent with the ``funnel-shaped potential" that follows 
from one gluon exchange and linear confinement. One gluon exchange implies
additional spin-dependent forces, specifically the contact 
spin-spin interaction (evident in the $J/\psi - \eta_c$ splitting) and  
spin-orbit and tensor forces that affect the fine structure of 
${\rm L} > 0$ multiplets. The agreement of the predicted splittings of the 
$\chi_{cJ}$ states with experiment (including the negative spin-orbit 
contribution of scalar confinement) has until recently been considered 
a clear success of this model, and is the strongest experimental evidence 
in favor of Lorentz scalar confinement.

The discovery of the X(3872), like the earlier reports of the 
$\D_{sJ}^*(2317)^+$ 
and 
$\D_{sJ}(2457)^+$, 
has called the accuracy of these models
into question. In both cases, narrow states have been reported 
at masses that are rather far from the predictions of quark potential 
models. Either these new states are not dominantly quarkonia, or 
we are seeing evidence of important additional forces that were not 
previously incorporated in the models.

\begin{table*}
\caption{Predicted and observed masses of 1D and 2P \ccbar states.}
\begin{tabular}{l|c|cccccc} \hline
State\phantom{,} 	& \, Expt. & \multicolumn{6}{c}{Theor.\phantom{X} } \\ 
	& 	&\phantom{,}GI\cite{God85}\phantom{,} 
                &\phantom{,}EF\cite{Eic81}\phantom{,} 
		&\phantom{,}FU\cite{Ful91}\phantom{,} 
		&GRR\cite{Gup86}
                &EFG\cite{Ebe03} 
		&ZVR\cite{Zen94}\\ 
\hline
 $1^3{\rm D}_3 $ & 	& 3849	& 3840 & 3884 & 3830	& 3815 & 3830 \\
 $1^3{\rm D}_2 $ & 	& 3838	& 3797 & 3871 & 3822	& 3813 & 3820 \\
 $1^3{\rm D}_1 $ & 3770 & 3819  & 3762 & 3840 & 3801	& 3798 & 3800 \\
 $1^1{\rm D}_2 $ & 	& 3837	& 3765 & 3872 & 3822	& 3811 & 3820 \\ \hline
 $2^3{\rm P}_2 $ & 	& 3979	&	&	&	& 3972 & 4020 \\
 $2^3{\rm P}_1 $ & 	& 3953	&	&	&	& 3929 & 3990 \\
 $2^3{\rm P}_0 $ & 	& 3916	&	&	&	& 3854 & 3940 \\
 $2^1{\rm P}_1 $ & 	& 3956	&	&	&	& 3945 & 3990 \\
\hline
\end{tabular}
\end{table*}

The most detailed predictions of the charmonium spectrum have come from
quark potential models. These models typically assume a color Coulomb 
plus linear confining interaction, which is augmented by the spin-dependent
forces that follow from one gluon exchange (OGE) and the confining interaction. 
These OGE terms are noncontroversial, and are the Breit-Fermi Hamiltonian
times a color factor; they consist of a contact spin-spin term, a spin-orbit
term and a smaller tensor interaction. The spin-dependent force that arises 
from confinement {\it is} rather controversial, as it depends on the assumed 
Lorentz structure of the confining interaction. The usual choice is 
scalar confinement, which gives an inverted spin-orbit term that partially
cancels the OGE term for small L. The alternative choice of vector confinement
(which was assumed in the Cornell model \cite{Eic76,Eic78,Eic80})
has a noninverted spin-orbit term, and unlike scalar confinement does not
give a good description of the splittings of the $\chi_{cJ}$ states.

The numerical mass predictions for the 1D and 2P  
\ccbar states given in Table~I are taken from several of these
potential models \cite{God85,Eic81,Ful91,Gup86,Ebe03,Zen94}; note
that most predict 1D states about $50-100$~MeV 
below the \X mass, and the 2P
states are predicted to lie above the \X by a similar amount.
The results are rather similar numerically because they differ
on relatively fine points such as 
relativizing quark motion, regularizing singular interactions, and the
choice of experimental input.
Clearly they all predict that the 1D \ccbar multiplet 
has a much smaller multiplet splitting than is implied by the \X and the
$\psi(3770)$. In this paper we tacitly assume that the potential model
wavefunctions are approximately correct for \ccbar states, and that the 
discrepancy in the spectrum is due to additional effects such as 
confinement spin-orbit terms 
or coupled-channel
effects, which shift the various \ccbar states by different amounts. 
The importance of these coupled-channel effects will be considered 
in future work.

Although the spectroscopy of charmonium states has been considered by 
many lattice gauge theory collaborations (for recent reviews see
Ref.\cite{Bal01,Bal03}), relatively few results have
been reported for the orbitally and radially excited 1D and 2P multiplets,
and these references quote rather large systematic and (for 2P) statistical
uncertainties, which at present imply an overall uncertainty of
roughly $\pm 100$~MeV \cite{Manke}.
The mean positions reported for the 1D \cite{Che01b,Lia02} 
and 2P \cite{Che01a,Che01b,Lia02,Aok02,Oka02}
multiplets are 
about 3.8~GeV and 4.0~GeV respectively, which are 
consistent with potential model estimates and with the 
experimental $1^3$D$_1$ state
$\psi(3770)$. Within the 1D multiplet there is some evidence 
from LGT that
the $3^{--}$ state lies above the $2^{--}$ and $2^{-+}$ 
states \cite{Lia02}. 
Lattice gauge theory predictions for these higher excitations are clearly
very important, and hopefully results with much smaller errors
will become available in future. Studies of the mass differences of
states within each multiplet would be especially interesting, and 
may be less sensitive to the large overall mass scale uncertainty.

\section{Radiative Transitions}

Radiative transitions can provide  
sensitive tests of
the spectroscopic assignments (angular quantum numbers) of
heavy-quark mesons. As an example, radiative transitions have been 
proposed \cite{God03,BAr03a,Col03}
as a means of determining the quantum numbers of the recently discovered 
$\D_{sJ}^*(2317)^+$ and $\D_{sJ}(2457)^+$ \cite{Aub03,Bes03,Kro03}.
In this section we calculate the E1 radiative widths that follow from
various \ccbar X(3872) assignments.

The partial width for an E1 radiative transition between
\ccbar states in the nonrelativistic quark model is given by 
\begin{widetext}
\be
\Gamma(
{\rm n}\, {}^{2{\S}+1}{\rm L}_{\J} 
\to 
{\rm n}'\, {}^{2{\S}'+1}{\rm L}'_{{\J}'}  
+ \gamma) 
 =  \frac{4}{3}\,  e_c^2 \, \alpha \,
\omega^3 \,   
C_{fi}\,
\delta_{{\S}{\S}'} \, 
|\,\langle 
{\rm n}'\, {}^{2{\S}'+1}{\rm L}'_{{\J}'} 
|
\; r \; 
|\, 
{\rm n}\, {}^{2{\S}+1}{\rm L}_{\J}  
\rangle\, |^2  
\ ,
\ee
\end{widetext}
(see for example Ref.\cite{Kwo88a}),
where 
$e_c= 2/3$ is the $c$-quark charge in units of $|e|$,
$\alpha$ is the fine-structure constant,
$\omega$ is the photon's energy, and the angular matrix element
$C_{fi}$ is given by
\be 
C_{fi}=\hbox{max}({\L},\; {\L}') (2{\J}' + 1)
\left\{ { {{\L}' \atop {\J}} {{\J}' \atop {\L}} {{\S} \atop 1}  } \right\}^2 .
\ee
For convenience the coefficients $\{ C_{fi}\} $ are listed 
in Tables~II and III.
The matrix elements 
$\langle {n'}^{2{\S}'+1}{\L}'_{{\J}'} |\; r \; 
| n^{2{\S}+1}{\L}_{\J}  \rangle$ 
are given in Tables~II and III, and were evaluated 
using the wavefunctions of Ref.\cite{God85}.
Relativistic corrections are implicitly included in these E1 
transitions through Siegert's theorem \cite{Sie37,McC83,Mox83}, 
by including spin-dependent interactions in the Hamiltonian used to 
calculate the meson masses and wavefunctions.   

We give two sets of predictions for these radiative widths.  In the 
first set (Table~II) we assume in all cases that the initial 
meson has the mass of the X(3872).  
While we appreciate that in some cases 
this is clearly an unlikely assignment,
such as the $1^3{\rm D}_1$ 
(normally identified with the $\psi(3770)$), 
we wish to consider decays of all conceivable 
X(3872) 
\ccbar 
assignments systematically, and 
will demonstrate that only a few 
possibilities are consistent with the existing \X data.  

In the second set of radiative width predictions (Table~III) 
we assume the \ccbar masses predicted by the Godfrey-Isgur model 
\cite{God85} where no obvious experimental candidate exists.  
This should generally give more reliable
predictions for the radiative widths of as yet unidentified \ccbar states, 
and will hopefully provide useful guidance for experimental searches 
for these states.

\begin{table*}
\caption{Radiative transitions in scenario 1: 
Predictions for the E1 transitions  
1D$\to$1P, 2P$\to$2S, 2P$\to$1S and 2P$\to$1D, 
assuming in all cases that the initial \ccbar state has a mass of
3872~MeV. The matrix elements were obtained using the wavefunctions 
of the Godfrey-Isgur model,
Ref.\cite{God85}.
Unless otherwise stated, the widths are given in keV 
and the final \ccbar masses are PDG values \cite{PDG02}.}
\begin{ruledtabular}
\begin{tabular}{l l c c c c c c } %\hline
Initial & Final & $M_f$ &  $\omega$ & 
	$\langle f| r | i \rangle $ & $C_{fi}$ & Width   \\
state X(3872) & state & (MeV) & (MeV) & (GeV$^{-1}$) &  & (keV) \\
\hline 
\\
$1^3{\rm D}_3 $ & $\chi_{c2}(1^3{\rm P}_2) \; \gamma$ 
	& \ 3556.2 & 303  & 2.762 & $\frac{2}{5}$ & 367 \\
\\
$1^3{\rm D}_2 $ & $\chi_{c2}(1^3{\rm P}_2) \; \gamma$ 
	& \ 3556.2 & 303 & 2.769 & $\frac{1}{10}$& 92 \\
	& $\chi_{c1}(1^3{\rm P}_1) \; \gamma$ 
	& \ 3510.5 & 345 & 2.588 & $\frac{3}{10}$ & 356 \\
\\
$1^3{\rm D}_1 $ & $\chi_{c2}(1^3{\rm P}_2) \; \gamma$ 
	& \ 3556.2 & 303 & 2.769 & $\frac{1}{90}$ & 10.2 \\
	& $\chi_{c1}(1^3{\rm P}_1) \; \gamma$ 
	& \  3510.5 & 345 & 2.598  & $\frac{1}{6}$ & 199 \\
	& $\chi_{c0}(1^3{\rm P}_0)\;  \gamma$ 
	& 3415\phantom{\footnotemark[1]} & 430 & 2.390  & $\frac{2}{9}$ & 437 \\
\\
$1^1{\rm D}_2 $ & $h_c(1^1{\rm P}_1) \; \gamma$ 
	& 3517\footnotemark[1] & 339
	& 2.627 & $\frac{2}{5}$ & 464 \\ 
\\
\hline
\\
$2^3{\rm P}_2 $ & $\psi'(2^3{\rm S}_1) \;  \gamma$ 
	& 3686\phantom{\footnotemark[1]} & 182 & 2.530 & $\frac{1}{3}$ & 55.2 \\
	& $J/\psi (1^3{\rm S}_1) \; \gamma$ 
	& 3097\phantom{\footnotemark[1]} & 697 & 0.276 & $\frac{1}{3}$ & 37.2 \\
	& $\psi''(1^3{\rm D}_1) \; \gamma$ 
	& 3770\phantom{\footnotemark[1]} & 101 & -2.031 & $\frac{1}{150}$ & 0.12 \\
	& $\psi_2(1^3{\rm D}_2) \; \gamma$ 
	& 3838\footnotemark[1] & 34 & -2.208 & $\frac{1}{10}$ & 0.08 \\
	& $\psi_3(1^3{\rm D}_3) \; \gamma$ 
	& 3849\footnotemark[1] & 23 & -2.375 & $\frac{14}{25}$ & 0.16 \\
\\
$2^3{\rm P}_1 $ & $\psi'(2^3{\rm S}_1) \; \gamma$ 
	& 3686\phantom{\footnotemark[1]} & 182 & 2.723 & $\frac{1}{3}$ & 63.9 \\
	& $J/\psi (1^3{\rm S}_1) \; \gamma$ 
	& 3097\phantom{\footnotemark[1]} & 697 & 0.150 & $\frac{1}{3}$ & 11.0 \\
	& $\psi''(1^3{\rm D}_1) \; \gamma$ 
	&  3770\phantom{\footnotemark[1]} & 101 & -2.244 & $\frac{1}{6}$ & 3.7 \\
	& $\psi_2(1^3{\rm D}_2) \; \gamma$ 
	& 3838\footnotemark[1] & 34 & -2.413 & $\frac{1}{2}$ & 0.49 \\
\\
$2^3{\rm P}_0 $ & $\psi'(2^3{\rm S}_1) \; \gamma$ 
	& 3686\phantom{\footnotemark[1]} & 182 & 2.899 & $\frac{1}{3}$ & 72.4 \\
	& $J/\psi (1^3{\rm S}_1) \; \gamma$ 
	& 3097\phantom{\footnotemark[1]} & 697 & -0.002 & $\frac{1}{3}$ & 1.5 eV \\
	& $\psi''(1^3{\rm D}_1) \; \gamma$ 
	& 3770\phantom{\footnotemark[1]} & 101 & -2.457 & $\frac{2}{3}$ & 17.8 \\
\\
$2^1{\rm P}_1 $ & $\eta_{c2}(1^1{\rm D}_2) \; \gamma$ 
	& 3837\footnotemark[1] & 35 & -2.395  & $\frac{2}{3}$ & 0.7 \\
	& $\eta_c' (2^1{\rm S}_0) \; \gamma$ 
	& 3638\footnotemark[2]& 227 & 2.303 & $\frac{1}{3}$ & 89 \\
	& $\eta_c (1^1{\rm S}_0) \; \gamma$ 
	& 2980\phantom{\footnotemark[1]} & 789 & 0.304 & $\frac{1}{3}$ & 65.4 \\
\end{tabular}
\end{ruledtabular}
\footnotetext[1]{Mass predicted by the Godfrey-Isgur model, Ref.\cite{God85}. 
The masses given in Ref.\cite{God85} were rounded to 
10~MeV; here we quote them to 1~MeV.}
\footnotetext[2]{Current world average, from Ref.\cite{Skw03}.}

\end{table*}

\begin{table*}
\caption{Radiative transitions in scenario 2: As in scenario 1, except that
unknown masses are taken from 
the Godfrey-Isgur model.}
\begin{ruledtabular}
\begin{tabular}{l l c c c c c } 
Initial & Final & $M_f$ &  $\omega$ & 
	$\langle f | r | i \rangle $ & $C_{fi}$ &  Width   \\
state  & state & (MeV) & (MeV) & (GeV$^{-1}$) &  & (keV) \\
\hline 
\\
$1^3{\rm D}_3(3849) $ 
& $\chi_{c2}(1^3{\rm P}_2) \; \gamma$ 
& $\ $ 3556.2\footnotemark[1] 
& 282 
& 2.762 
& $\frac{2}{5}$ 
& 295 \\
\\
$1^3{\rm D}_2(3838) $ 
& $\chi_{c2}(1^3{\rm P}_2) \; \gamma$ 
& $\ $ 3556.2\footnotemark[1] 
& 271 
& 2.769 
& $\frac{1}{10}$ 
& 66 \\
& $\chi_{c1}(1^3{\rm P}_1) \; \gamma$  
& $\ $ 3510.5\footnotemark[1] 
& 314 
& 2.588 
& $\frac{3}{10}$ 
& 268 \\
\\
$1^3{\rm D}_1(3770)\footnotemark[1] $ 
& $\chi_{c2}(1^3{\rm P}_2) \; \gamma$ 
& $\ $ 3556.2\footnotemark[1] 
& 208 
& 2.769 
& $\frac{1}{90}$ 
& 3.3 \\
	
& $\chi_{c1}(1^3{\rm P}_1) \; \gamma$   
& $\ $ 3510.5\footnotemark[1] 
& 251 
& 2.598 
& $\frac{1}{6}$ 
& 77 \\
& $\chi_{c0}(1^3{\rm P}_0) \; \gamma$ 
& 3415\footnotemark[1] 
& 338 
& 2.390 
& $\frac{2}{9}$ 
& 213 \\
\\
$1^1{\rm D}_2(3837) $ 
& $h_c(1^1{\rm P}_1) \; \gamma$ 
& 3517\phantom{\footnotemark[1]} 
& 307 	
& 2.627 
& $\frac{2}{5}$ 
& 344 \\ 
\\
\hline
\\
$2^3{\rm P}_2(3979) $ 
& $\psi'(2^3{\rm S}_1) \; \gamma$ 
& 3686\footnotemark[1] 
& 282 
& 2.530 
& $\frac{1}{3}$ 
& 207 \\
	
& $J/\psi (1^3{\rm S}_1) \; \gamma$ 
& 3097\footnotemark[1] 
& 784 
& 0.276 
& $\frac{1}{3}$ 
& 53 \\
& $\psi_3(1^3{\rm D}_3) \; \gamma$ 
& 3849\phantom{\footnotemark[1]} 
& 128
& -2.375 
& $\frac{14}{25}$  
& 29 \\
& $\psi_2(1^3{\rm D}_2) \; \gamma$ 
& 3838\phantom{\footnotemark[1]} 
& 139 
& -2.208 
& $\frac{1}{10}$  
& 5.6 \\
& $\psi''(1^3{\rm D}_1) \; \gamma$ 
& 3770\footnotemark[1] 
& 204
& -2.031 
& $\frac{1}{150}$  
& 1.0 \\
\\
$2^3{\rm P}_1(3953) $ 
& $\psi'(2^3{\rm S}_1) \; \gamma$ 
& 3686\footnotemark[1] 
& 258 
& 2.723 
& $\frac{1}{3}$ 
& 184 \\
& $J/\psi (1^3{\rm S}_1) \; \gamma$ 
& 3097\footnotemark[1] 
& 763 
& 0.150 
& $\frac{1}{3}$ 
& 14.4 \\
& $\psi_2(1^3{\rm D}_2) \; \gamma$   
& 3838\phantom{\footnotemark[1]} 
& 113 
& -2.413 
& $\frac{1}{2}$ 
& 18.3 \\
& $\psi''(1^3{\rm D}_1) \; \gamma$   
& 3770\footnotemark[1] 
& 179 
& -2.244 
& $\frac{1}{6}$ 
& 20.7 \\
\\
$2^3{\rm P}_0(3916) $ 
& $\psi'(2^3{\rm S}_1) \; \gamma$ 
& 3686\footnotemark[1] 
& 223 
& 2.899
& $\frac{1}{3}$ 
& 135 \\
& $J/\psi (1^3{\rm S}_1) \; \gamma$ 
& 3097\footnotemark[1] 
& 733 
& -0.002 
& $\frac{1}{3}$ 
& 1.6 eV \\
& $\psi''(1^3{\rm D}_1) \; \gamma$ 
& 3770\footnotemark[1] 
& 143 
& -2.457 
& $\frac{2}{3}$ 
& 51.2 \\
\\
$2^1{\rm P}_1(3956) $ 
& $\eta_{c2}(1^1{\rm D}_2) \gamma$ 
& 3837\phantom{\footnotemark[1]} 
& 117 
& -2.395 
& $\frac{2}{3}$  
& 26.6 \\
& $\eta_{c}' (2^1{\rm S}_0) \; \gamma$ 
& 3638\footnotemark[2] 
& 305 
& 2.303 
& $\frac{1}{3}$ 
& 217 \\
& $\eta_c (1^1{\rm S}_0)\; \gamma$ 
& 2980\footnotemark[1] 
& 856 
& 0.304 
& $\frac{1}{3}$ 
& 83 \\
\end{tabular}
\end{ruledtabular}
\footnotetext[1]{Experimental PDG mass \cite{PDG02}.}
\footnotetext[2]{Current world average, from Ref.\cite{Skw03}.}

\end{table*}

\section{Strong Decays}

Strong decays provide crucial tests of the nature of the X(3872),
through the total width and relative branching fractions.
We consider three types of strong decays,

\vskip 0.5cm
\noindent
1) Zweig-allowed open-charm decays, 
$(c\bar c) \to (c\bar q) + (q\bar c)$ ($q=u,d,s)$, 

\vskip 0.5cm
\noindent
2) \ccbar annihilation,  
$c\bar{c}\to gg, \; ggg, \; q\bar{q} g , 
\ldots$  

\vskip 0.5cm
\noindent
3) closed-flavor hadronic transitions, such as
$(c\bar{c})\to J/\psi\, \pi\pi, \eta_c\, \pi\pi,
J/\psi\,\eta, \eta_c\,\eta, \ldots$

\vskip 0.5cm
We estimate the Zweig allowed decays using the $^3$P$_0$ decay model.  
The history of this model and related strong decay models has been 
reviewed recently by Barnes \cite{Bar03b}; details of the approach
may be found in the extensive literature (see for example
Ackleh {\it et al.} \cite{Ack96} and Blundell and Godfrey \cite{Blu96}). 
The $^3$P$_0$ strong decay amplitudes are
given by a dimensionless pair production amplitude $\gamma$ 
times a convolution integral of the three meson wavefunctions. 
Based on our experience with light meson decays we set $\gamma = 0.4$.
We assume SHO wavefunctions for the three mesons,
with a universal Gaussian width parameter of $\beta=0.5$~GeV;
this is a rough average of $\beta$ values that give maximum overlap
with nonrelativistic Coulomb plus linear wavefunctions as well as 
Godfrey-Isgur wavefunctions. 
We also generalized the $^3$P$_0$ decay overlap integrals of 
Ackleh {\it et al.} \cite{Ack96} 
to accommodate different quark and antiquark masses in the final 
mesons. The single new parameter required here 
is the heavy-light quark mass ratio
$r = m_c/m_q$, which we take to be $1.5/0.33$ for $u,d$ and 
$1.5/0.55$ for $s$.
Our results for the partial widths of these open-flavor modes 
are given in Table~IV (with all initial masses set to 3872~MeV) 
and Table~V (with all unknown masses set to the Godfrey-Isgur values). 

Note that the $1^3\D_2$ $\psi_2$ and the $1^1\D_2$ $\eta_{2c}$ cannot
decay to $\D{\bar {\rm D}}$ due to parity conservation, and since they are
below the next open flavor threshold (DD$^*$) they are expected
to be narrow. The narrowness of the $1^3\D_3$ $\psi_3$ in contrast
is due to suppression by the $\D{\bar {\rm D}}$ 
F-wave angular momentum barrier. 

Experience with light and strange meson strong decays suggests that 
these partial widths should be accurate to perhaps a factor of two 
(given the correct masses); 
the predicted width of the $\psi(3770)$ (in Table~V), for example, is
$\Gamma(1^3\D_1(3770) \to \D{\bar {\rm D}} ) = 42.8$~MeV,
whereas the PDG experimental average is 
$\Gamma^{tot}_{\psi(3770)} = 25.3 \pm 2.9 $~MeV \cite{PDG02}.

Annihilation decays into gluons and light quarks
make significant contributions to the total widths of some
\ccbar resonances. These decay rates 
been studied extensively using pQCD methods
\cite{App75,DeR75,Cha75,Bar76a,Bar76b,Nov78,Bar79,
Kwo88b,Ber91,Ack92a,Ack92b};  
the relevant formulas are summarized in Ref.\cite{Kwo88b}.  
Expressions for decay widths relevant to the 1D and 2P \ccbar states 
are:
\begin{eqnarray} 
&\Gamma(^3\D_J \to ggg)  = &\frac{10\alpha_s^3}{9\pi}\, C_J\, 
	\frac{|R_{\D}''(0)|^2}{m_Q^6} \ln (4 m_Q\langle r \rangle ) \\
&\Gamma(^1\D_2 \to gg)  = &\frac{2\alpha_s^2}{3}\, 
	\frac{|R_{\D}''(0)|^2}{m_Q^6} \\
&\Gamma(^3\P_2 \to gg)  = &\frac{8\alpha_s^2}{5}\, 
	\frac{|R_{\P}'(0)|^2}{m_Q^4} \\
&\Gamma(^3\P_1 \to q\bar{q}g)  = &\frac{8 n_f \alpha_s^3}{9\pi}\, 
	\frac{|R_{\P}'(0)|^2}{ m_Q^4} \ln (m_Q\langle r \rangle ) \\
&\Gamma(^1\P_1 \to ggg)  = &\frac{20\alpha_s^3}{9\pi}\, 
	\frac{|R_{\P}'(0)|^2}{m_Q^4} \ln (m_Q\langle r \rangle ) \\
&\Gamma(^1\P_1 \to gg  \gamma )  = &\frac{36}{5} e_q^2\, {\alpha \over 
	\alpha_s}\, \Gamma(^1\P_1 \to ggg) \\
&\Gamma(^3\P_0 \to gg)  = &6 \alpha_s^2\, 
	\frac{|R_{\P}'(0)|^2}{m_Q^4} \\ 
\end{eqnarray}
where $C_J = \frac{76}{9}$, 1, 4 for J = 1, 2, 3, and the 
number of light quarks is taken to be $n_f=3$. To obtain our numerical 
results for these partial widths we assumed $m_c=1.628$~GeV, 
$\alpha_s\approx 0.23$ (with some weak mass dependence),
and used the wavefunctions of Ref.\cite{God85}.

Considerable uncertainties arise in these expressions from the 
model-dependence of the wavefunctions and possible relativistic 
and QCD radiative corrections (see for example the discussion
in Ref.\cite{God85}). As one example of a likely inaccuracy, 
the contact approximation for 
$c\bar c\; (1^1\D_2) \to gg$ given above 
has been checked numerically, and overestimates the rate 
found with a full $c$-quark propagator by 
about two orders of magnitude \cite{Ack92a}. 
Other problems are that 
the logarithm evident in some of these formulas is evaluated at
a rather arbitrarily chosen scale, 
and that the pQCD radiative corrections to these processes
are often found to be large, but are prescription dependent and so are
numerically unreliable. Thus, we regard these formulas as rough estimates
of the partial widths for these annihilation processes rather than accurate
predictions, and they certainly merit more theoretical effort in the future.  
The numerical partial widths we find for these annihilation 
processes are given in Tables~IV and V.

The final strong decays we consider are closed-flavor hadronic 
transitions of the type $(c\bar{c})\to (c\bar{c}) + \pi\pi (\eta)$. 
There have been many theoretical estimates of these and related 
transitions 
\cite{Yan80,Kua81,Kua88,Kua90,Vol80,Vol86,Iof80,Nov81,Vol03a,Bel87,
Mox88,Eic94,Bar03a}.  
Here we employ the multipole expansion of color gauge fields 
developed by Yan and collaborators \cite{Yan80,Kua81,Kua88,Kua90}
together with the Wigner-Eckart theorem 
to estimate the E1-E1 transition rates \cite{Yan80};
the relevant expressions are summarized by Eichten and Quigg \cite{Eic94}. 
The recent BES measurement of 
the B$(\psi(3770) \to J/\psi \, \pi^+\pi^-)=(0.59\pm 0.26 
\pm 0.16)\% $ \cite{Bai03} is used
as input for the \ccbar transitions 
of the type $(c\bar c)_{\D} \to (c\bar c)_{\S} \pi \pi$.  
One should be cautious about this result and the predictions we derive 
from it as CLEO-c has presented the smaller preliminary limit of 
B$(\psi(3770) \to J/\psi\, \pi^+\pi^-)<0.26\%$ at 90\% C.L.
\cite{Skw03}.
Furthermore, rescaling the $A_2(2,0)$ $b\bar{b}$ amplitude needed for 
the $\D\to \S$ transitions gives 
$\Gamma(\psi(3770) \to J/\psi\, \pi^+\pi^-)\simeq 58$~keV, which is 
consistent with the CLEO-c result but is 
is about a factor of 2 smaller than the BES measurement.  
The hadronic transition rates, based on the BES measurement,
are summarized with the other strong decays in Tables~IV and V. 
We do not include decays of the type 
$2^{3,1}\P_J\to 1^{3,1}\P_{J'}$, as they are expected to be 
small compared to the decays considered here.  
Similarly, transitions 
with $\eta$ and $\pi^o$ in the final state are also possible but are 
expected to have much smaller partial widths than the decays that we have 
included.

\section{Discussion of X(3872) $c\bar c$ assignments}

A summary of the strong and electromagnetic  
partial widths predicted for each
1D and 2P \ccbar assignment for the \X is given in Table~IV.
The initial mass in all cases is taken to be 3872~MeV.

One may immediately eliminate the 
$2^3$P$_2$,
$2^3$P$_0$ 
and the
``straw dog" assignment 
$^3$D$_1$,
due to the large theoretical total widths.
The total width of a $^3$D$_1(3872)$ state is predicted to be about
two orders of magnitude larger than the experimental
limit of 2.3~MeV (95\% C.L.) for the X(3872), and 
$2^3$P$_2$ and
$2^3$P$_0$ states at 3872~MeV would have 
strong widths 
an order of magnitude larger than the experimental limit.
(We note that the process $2^3$P$_0 \to $ D$\bar{\rm D}$
is accidentally near a node in the decay amplitude, which 
gives a suppressed rate for this S-wave decay. This may be an artifact 
of the decay model. In any case annihilation decays should
insure that the $2^3$P$_0$ \ccbar is not a narrow state.)

{\it A priori} the most plausible \ccbar assignments for the \X 
are
$1^3$D$_2$
and
$1^1$D$_2$. Since the mode 
D$\bar{\rm D}$ is forbidden, 
these states have no allowed open-charm decay mode, 
and must decay instead 
through the weaker short-distance \ccbar annihilation 
processes, radiative decays, and closed-flavor hadronic transitions. 
We find that the these decays lead to theoretical total widths 
of about 1~MeV for both these states. 

These $2^-$ states should both have quite large E1 radiative 
branching fractions, in total $\approx 50\%$, and the final states are 
very characteristic. 
The spin-triplet $1^3$D$_2$ will decay into $\chi_{c2}\gamma$ and 
$\chi_{c1}\gamma$
with a relative branching fraction of about $1:4$, whereas
the spin-singlet $1^1$D$_2$ will decay into $h_c\gamma$, where
the $h_c$ is the as yet 
unidentified spin-singlet P-wave state. 
Confirmation of a $1^1$D$_2$ \ccbar \X assignment 
may therefore require the identification of the 
problematic $1^1$P$_1$ \ccbar state.

We find that the current Belle limit 
on the radiative decay of the X(3872),
\begin{equation}
{{\B}(\X\to \chi_{c1} \gamma)\over 
{\B}(\X\to J/\psi \, \pi^+\pi^- )} \ < \ 0.89, \ \ 90\% \ {\rm C.L.}
\end{equation}
is only marginally a problem for the $1^3\D_2$ \ccbar assignment,
due to our larger scale (relative to Ref.\cite{Eic02}) and 
significant uncertainty in the
$J/\psi\,\pi\pi$ branching fraction (see Table~IV).  However, the recent 
CLEO-c result would pose a problem for the prediction and we eagerly 
await more precise data from these experiments. 
With somewhat better experimental statistics we anticipate that the 
$\chi_{c1} \gamma$
and
$\chi_{c2} \gamma$
modes will both be evident, if the X(3872) is indeed a 
$1^3\D_2$ \ccbar state.

Although the $1^3$D$_3$ \ccbar state does have an open-charm 
decay mode (D${\bar {\rm D}}$),
we find that the centrifugal barrier actually implies a small total width of
only a few MeV; given the uncertainties in the $^3$P$_0$ decay model, this state
should also be considered a viable \X candidate. The $1^3$D$_3$ assignment can
also be tested by studying radiative decays; this state is predicted
to have an 8\% branching fraction to $\chi_{c2}\gamma$, 
but $\chi_{c1}\gamma$ in contrast is M2, and will have a much smaller 
partial width.  Thus the $\chi_{c1}\gamma$
and $\chi_{c2}\gamma$ decay modes can be used to distinguish between 
$1^3\D_2$ and $1^3\D_3$.

The 
$2^3$P$_1$
and
$2^1$P$_1$ 
states if at 3872~MeV would have total widths of about
1-2~MeV, also consistent with the \X experimental limit.
These states are notable in that they should {\it not} be
clearly evident in radiative transitions; E1 branching fractions
of only a few percent are expected, and unlike the E1 decays 
of D-wave charmonia, these 2P states do not populate the 
modes $\chi_c\gamma$ or $h_c\gamma$; instead an initial
$2^3$P$_1$
or
$2^1$P$_1$ 
leads to
$(J/\psi,\psi') \gamma$ or $(\eta_c,\eta_c') \gamma$ 
respectively.  Problems with these 2P assignments are that
we do not expect the $J/\psi\, \pi\pi$ 
final state to be prominent, and the predicted masses are roughly 
100~MeV higher than the X(3872).

The search for a 
$J/\psi\, \pi^o \pi^o$ mode is a very important 
experimental task. If the \X is indeed a \ccbar state,
the presence or absence of this mode will select C-parity
$(-)$ or $(+)$ respectively \cite{Voloshin}. 
Decays to $J/\psi\, \pi^o\pi^o$ imply that the initial state has C $= (-)$,
whereas if the 
decay proceeds through an isospin-violating transition to 
$J/\psi\, \rho^o$ followed by $\rho^o\to \pi^+\pi^-$, the
initial state has C $ = (+)$.  In the former case the 
$J/\psi\, \pi^o \pi^o$ mode should have
approximately 1/2 the branching fraction of $J/\psi\, \pi^+\pi^-$
(expected for I=0).  In contrast, the $\rho^o$ only decays to 
charged pions.  The observation of this
state in a $J/\psi \,\pi\pi$ mode and past experience with 
dipion decays suggests C $ = (-)$, but this should be 
checked through a search for $J/\psi\, \pi^o \pi^o$.
If a $J/\psi\, \pi^o \pi^o$ decay mode is confirmed with this strength,
we are then left with the C$ = (-)$ \ccbar candidates
$1^3$D$_3$, 
$1^3$D$_2$ 
and
$2^1$P$_1$. 
Conversely, if there is no significant $J/\psi\, \pi^o \pi^o$ mode
relative to $J/\psi\, \pi^+\pi^-$, the \X would presumably be
C $ = (+)$, with \ccbar candidates
$1^1$D$_2$ 
and
$2^3$P$_1$. 
Studies of radiative decays can be used to test the remaining \ccbar 
possibilities once the C-parity is established.  
We note in passing that the pion invariant mass distribution has also been 
advocated as a discriminator between these assignments \cite{Yan80,Pak03}.

If we use the mass predictions of the Godfrey-Isgur model 
(instead of the X(3872) mass) 
to calculate the properties of 
1D and 2P \ccbar states (Table~V),
we find 
that all of the 2P states are rather broad, making them more 
difficult to observe in B decay.  In contrast all the 1D states remain 
relatively narrow, since the predicted Godfrey-Isgur masses are below
the X(3872) mass.  We therefore expect that all members of the 1D 
multiplet will be observable in B meson decays, independent of the nature 
of the X(3872).

\section{Non-$c\bar c$ assignments: DD$^*$ Molecule}

The fact that the reported \X mass and the D$^o$D$^{*o}$ threshold 
are equal to within the current errors of about 1~MeV has led to
speculations that this state might actually be a weakly bound 
DD$^*$ molecule, perhaps dominantly 
D$^o$D$^{*o}$ \cite{Tor03,Clo03b,Vol03b,Pak03,Won03,Bra03,Swa03}. 
The possibility of charm meson molecules has been discussed in
several earlier references, especially regarding the 
$\psi(4040)$ as a D$^*$D$^*$ candidate
\cite{Nov78,Vol76,DeR76}.

DD$^{*}$ molecule assignments can be distinguished from 
\ccbar through quantum numbers and decay modes. Since a molecular
state would most likely be an S-wave, J$^{\P} = 1^+$ is expected.
Either C-parity is possible {\it a priori}, and attractive forces
do arise in each C-parity channel, due to strong virtual decay 
couplings to the (theoretically) 
higher-mass C $ = (+)$ $2^3\P_1$ and C $ = (-)$ $2^1\P_1$ \ccbar states. 

Assuming binding from one pion exchange forces, T\"ornqvist \cite{Tor03} 
argues that C $ = (+)$, and in addition to the S-wave
J$^{\rm PC} = 1^{++}$ state the P-wave combination with 
J$^{\rm PC} = 0^{-+}$ should also be bound. Since a C $ = (+)$ state
cannot decay to $J/\psi\,(\pi\pi)_S$, T\"ornqvist suggests that the 
observed $J/\psi\,\pi^+\pi^-$ final state may be due to a
$J/\psi\,\rho^o$ decay, allowed by
isospin mixing in the initial state \cite{Rosner}. (A dominantly
D$^o$D$^{*o}$ molecule for example has I=0 and I=1 components with
comparable weight \cite{Clo03b}.)
Swanson \cite{Swa03} finds that attraction from pion exchange alone 
is not sufficient to form a DD$^*$ molecular bound state,
but that a J$^{\rm PC} = 1^{++}$ 
bound state does form when short-ranged quark-gluon forces
are included as well.

In a hypothetical very weakly bound, dominantly D$^o$D$^{*o}$ molecule, 
one would expect the decays and partial widths
to be essentially those of the D$^{*o}$. This implies dominant decay
modes D$^o$D$^o\pi^o$ and D$^o$D$^o\gamma$, 
in an approximately 
$1.5:1$ ratio, and a total width equal to that of the D$^{*o}$,
which is theoretically $\approx 50$~keV. Swanson \cite{Swa03} in contrast 
finds that internal 
rescatter is important in the D$^o$D$^{*o}$ bound system, which leads 
instead to dominant 
$J/\psi\,\rho^o$ and $J/\psi\,\omega$ 
decay modes, giving a total width of 
$\approx 2$~MeV. This is essentially equal to the current experimental 
limit. A search for the $J/\psi\,\omega$ mode would be an important 
test of this molecule decay model.

There appears to be general agreement  
that the radiative transitions of a weakly bound molecule to any 
$(c{\bar c})\gamma$ channel should be highly suppressed, so 
establishing limits on 
these radiative partial widths would also provide useful 
tests of DD$^*$ molecule models.

One should note that mixing between the DD$^*$ and
\ccbar basis states will certainly be present at some level, 
so even in a dominantly molecular DD$^*$ state, suppressed 
transitions from the \ccbar component of the \X to $(c{\bar c})\gamma$ 
will occur. The observed radiative partial widths relative to predictions 
for pure \ccbar states can be used to quantify the  
$(c{\bar c}) \leftrightarrow $DD$^*$ mixing.

\section{Non-$c\bar c$ assignments: Charmonium Hybrid}

Charmonium hybrid states have been predicted to have masses 
in the range of 4.0 
to 4.4~GeV, with the higher value preferred by recent LGT studies.  
The flux-tube decay model argues that these states will be narrow
if they lie below the S+P open-charm threshold DD$^*_J$, 
and hence will have a
relatively large branching fraction to 
$J/\psi\, \pi\pi$. (Of course the large branching fraction reported for
$\pi_1(1600) \to \rho\pi$ argues against dominance by high-mass
S+P decay modes.) 
Charmonium hybrids are also expected to have
relatively small radiative widths. Although the 
reported properties of the X(3872) are consistent with these  
expectations for $2^{+-}$ and $0^{+-}$ hybrids, 
the large discrepancy with the predicted LGT mass of 4.4~GeV 
makes this assignment appear unlikely. 
In addition, a recent lattice study finds that some hybrid closed-flavor 
decays have surprisingly large partial widths \cite{McN02},
which may also argue against a hybrid assignment for the X(3872).
  
\section{Conclusions}

In this paper we have considered all 
possible 1D and 2P \ccbar assignments for the recently
discovered X(3872), 
since these are the only \ccbar states expected near the mass of the 
X(3872).
In particular we evaluated the strong and electromagnetic
partial widths of all states in these multiplets, and compared the 
results to our current knowledge of the X(3872).

Assuming a mass of 3872~MeV, the large predicted total widths
eliminate the $1^3\D_1$, $2^3\P_2$ and $2^3\P_0$ as candidates,
leaving the 
$1^3\D_3$, $1^3\D_2$, $1^1\D_2$, $2^3\P_1$ and $2^1\P_1$. 
A search for the mode $J/\psi\, \pi^o\pi^o$ will be important to 
discriminate between these remaining possibilities. 
The observation of a $J/\psi\, \pi^o\pi^o$ mode with a relative 
$J/\psi\, \pi^+\pi^-$ branching fraction of approximately 1~:~2
indicates a C $ = (-)$ state, and would
restrict the plausible \ccbar 
\X assignments to $1^3\D_3$, $1^3\D_2$, and $2^1\P_1$.
A limit on $J/\psi\, \pi^o\pi^o$ well below this 1~:~2 ratio 
would imply C~$ = (+)$, leaving $1^1\D_2$ and $2^3\P_1$
as possible assignments. A unique assignment can then be established 
through studies of the final states populated in \X radiative transitions.  
The observation of a $J/\psi\, \pi^o\pi^o$ signal with a strength 
comparable to $J/\psi\, \pi^+\pi^-$ but significantly different from the 
1~:~2 ratio would indicate that the initial state is not an I-spin 
eigenstate; depending on the value of this ratio,
this might support a mixed-isospin DD$^*$ molecule interpretation.

Radiative transitions have previously been advocated as important tests 
of the nature of the \X because the estimated rates vary widely for different
types of initial states, and the radiative partial widths between 
pure \ccbar basis states can be calculated with reasonable accuracy 
(of perhaps 30\%). 
For pure 1D \ccbar assignments for the X(3872), we find that the 
relative branching fractions to the modes 
$\chi_{c1}\gamma$, 
$\chi_{c2}\gamma$
and
$h_{c}\gamma$
depend strongly on the initial state, and can be used to distinguish
between $1^3\D_3$, $1^3\D_2$ and $1^1\D_2$.
We noted however that as the \X is essentially 
degenerate with the DD$^*$ threshold, we expect a significant DD$^*$ 
component in X(3872), even if it is dominantly a \ccbar state.
Thus if $c\bar{c} \leftrightarrow {\rm DD}^*$ mixing is significant,
we would expect radiative transitions to $(c\bar c)\gamma$ to be observed, 
but with
partial widths that are suppressed relative to the predictions
of the \ccbar quark model. Similarly DD$^*$-molecule decay modes such as
D$^o$D$^o\gamma$, D$^o$D$^o\pi^o$, 
$J/\psi\,\rho^o$ and $J/\psi\,\omega$ 
should also be present in a mixed 
$c\bar{c} - {\rm DD}^*$ state, but at a suppressed rate relative
to the partial width expected from a dominantly DD$^*$ molecular
bound state.

As an interesting final observation, we expect 
the $1^3\D_3$ $\psi_3$ to be rather 
narrow, and to have significant branching fractions to $J/\psi\, \pi\pi$ 
and $\chi_{c2}\gamma$. This suggests that the $\psi_3$ 
should be observable in B decay.  
The observation of all the members of the 1D \ccbar multiplet
would contribute very useful information to the study of 
spin-dependent forces in heavy quarkonia.

\acknowledgments

We would like to acknowledge useful discussions with F.E.Close and 
E.S.Swanson, and
communications from S.Conetti, G.Bali, T.Manke, C.Michael, A.Petrov, 
J.Rosner, and T.Skwarnicki.
SG would also like to thanks the organizers of the QWG2 workshop 
for providing a stimulating environment for discussions of the 
X(3872).
This research was supported in part by the 
the Natural Sciences and Engineering Research Council of Canada, 
the U.S. National Science
Foundation through grant NSF-PHY-0244786 at the University of Tennessee, 
and the U.S. Department of Energy under contract DE-AC05-00OR22725 at
Oak Ridge National Laboratory (ORNL).

%\begin{references}

\vfill\eject

\begin{table}
\caption{Partial widths and branching fractions
for strong and electromagnetic transitions in scenario 1:
We assume in all cases that the initial \ccbar 
state has a mass of 3872~MeV.  
Details of the calculations are given in the text.
\label{tab:sum1}}
\begin{center}
\begin{tabular}{l l l r } \hline \hline
Initial \phantom{www}        & Final            	& Width  	& B.F. \\
 state          & state            	& (MeV)  	& \ \ (\%)  \\ \hline
$1^3{\rm D}_3$	& DD	   	& 4.04	 	& 84.2	\\
		& $ggg$		   	& 0.18	 	& 3.8	\\
		& $ J/\psi \pi\pi $ 	& $0.21\pm 0.11$& 4.4	\\
		& $\chi_{c2}(1^3\P_2) \gamma$ & 0.37	& 7.7	\\
		& Total			& 4.80 		& 100  \\
\hline
$1^3{\rm D}_2$	& $ggg$		  	&  0.08		& 10.8 \\
		& $ J/\psi \pi\pi $ 	& $0.21\pm 0.11$& 28.4 \\
		& $\chi_{c2}(1^3\P_2) \gamma$ & 0.09	& 12.2 \\
		& $\chi_{c1}(1^3\P_1) \gamma$ & 0.36	& 48.6 \\
		& Total			& 0.74 		& 100  \\
\hline
$1^3{\rm D}_1$	& DD 		& 184		& 98.9 \\
		& $ggg$		   	& 1.15 		& 0.6 \\
		& $ J/\psi \pi\pi $ 	& $0.21\pm 0.11$& 0.1 \\
		& $\chi_{c1}(1^3\P_1) \gamma$ & 0.20	& 0.1 \\
		& $\chi_{c0}(1^3\P_0) \gamma$ & 0.44	& 0.2 \\
		& Total			& 186 		& 100 \\
\hline
$1^1{\rm D}_2$	& $gg$		 	& 0.19 		& 22.1 \\
		& $ \eta_c \pi\pi$	&$0.21\pm 0.11$	& 24.4 \\ 
		& $h_c(1^1\P_1) \gamma$ 	& 0.46		& 53.5 \\
		& Total			& 0.86 		& 100 \\
\hline
\hline
$2^3{\rm P}_2$	& DD		& 21.1		& 82.4 \\
		& $gg$		   	& 4.4 		& 17.2 \\
		& $\psi'(2^3\S_1) \gamma$ & 0.06		& 0.2 \\
		& $J/\psi(1^3\S_1) \gamma$ & 0.04	& 0.2 \\
		& Total			& 25.6		& 100 \\
\hline
$2^3{\rm P}_1$	& $q\bar{q}g$		& 1.65 		& 95.9 \\
		& $\psi'(2^3\S_1) \gamma$ & 0.06		& 3.5 \\
		& $J/\psi(1^3S_1) \gamma$ & 0.01	& 0.6 \\
		& Total			& 1.72 		& 100 \\
\hline
$2^3{\rm P}_0$	& DD		& 13.7 (see text) 	& 24.6 \\
		& $gg$		   	& 42. 		& 75.3 \\
		& $\psi'(2^3\S_1) \gamma$ & 0.07		& 0.1 \\
		& $\psi''(1^3\D_1) \gamma$ & 0.02 	& $4\times 10^{-2}$\\
		& Total			& 55.8		& 100 \\
\hline
$2^1{\rm P}_1$	& $ggg$		   	& 1.29 		& 81.6 \\
		& $gg\gamma$		& 0.13 		& 8.2 \\
		& $\eta_c'(2^1\S_0) \gamma$ & 0.09	& 5.7 \\
		& $\eta_c(1^1\S_0) \gamma$ & 0.07	& 4.4 \\
		& Total			& 1.58		& 100 \\
\hline \hline
\end{tabular}
\end{center}

\end{table}

\vfill\eject

\begin{table}
\caption{As in Table.IV, except that
unknown masses are taken from the Godfrey-Isgur model.
\label{tab:sum2}}
\begin{center}
\begin{tabular}{l l l r } \hline \hline
Initial         & Final            	& Width  	& B.F. \\
 state          & state            	& (MeV)  	& \ \ (\%) \\ \hline
$1^3\D_3(3849)$\phantom{www}	& DD	   	& 2.27	 	& 76.7 \\
		& $ggg$		   	& 0.18	 	& 6.1 \\
		& $ J/\psi \pi\pi $ 	& $0.21\pm 0.11$& 7.1 \\
		& $\chi_{c2}(1^3\P_2) \gamma$ & 0.30	& 10.1 \\
		& Total			& 2.96 		& 100 \\
\hline 
$1^3\D_2(3838)$	& $ggg$		  	&  0.08		& 12.7 \\
		& $ J/\psi \pi\pi $ 	& $0.21\pm 0.11$& 33.3 \\
		& $\chi_{c2}(1^3\P_2) \gamma$ & 0.07	& 11.1 \\
		& $\chi_{c1}(1^3\P_1) \gamma$ & 0.27	& 42.9 \\
		& Total			& 0.63 		& 100 \\
\hline 
$1^3\D_1(3770)$	& DD 		& 42.8  	& 96.4 \\
		& $ggg$		   	& 1.15 		& 2.6 \\
		& $ J/\psi \pi\pi $ 	& $0.21\pm 0.11$& 0.5 \\
		& $\chi_{c1}(1^3\P_1) \gamma$ & 0.08	& 0.2 \\
		& $\chi_{c0}(1^3\P_0) \gamma$ & 0.21	& 0.5 \\
		& Total			& 44.4 		& 100 \\
\hline 
$1^1\D_2(3837)$	& $gg$		 	& 0.19 		& 25.7 \\
		& $ \eta_c \pi\pi$	&$0.21\pm 0.11$	& 28.4 \\ 
		& $h_c(1^1\P_1) \gamma$ 	& 0.34		& 45.9 \\
		& Total			& 0.74 		& 100 \\
\hline
\hline
$2^3\P_2(3979)$	& DD		& 42.4		& 46.8 \\
		& DD$^*$		& 42.5		& 46.9 \\
		& $\D_s \D_s$	& 1.03		& 1.1 \\
		& $gg$		   	& 4.4 		& 4.9 \\
		& $\psi'(2^3\S_1) \gamma$ & 0.21		& 0.2 \\
		& $J/\psi(1^3\S_1) \gamma$ & 0.05 & $6\times 10^{-2}$\\
		& $\psi_3(1^3\D_3) \gamma$ & 0.03 & $3\times 10^{-2}$\\
		& Total			& 90.6 		& 100 \\
\hline 
$2^3\P_1(3953)$	& DD$^*$		& 118.		& 98.4 \\
		& $q \bar{q} g$		& 1.65 		& 1.4 \\
		& $\psi'(2^3\S_1) \gamma$ & 0.18		& 0.2\\
		& $J/\psi(1^3\S_1) \gamma$ & 0.01 & $8\times 10^{-3}$\\
		& $\psi_2(1^3\D_2) \gamma$ & 0.02 & $2\times 10^{-2}$\\
		& $\psi_1(1^3\D_1) \gamma$ & 0.02 & $2\times 10^{-2}$\\
		& Total			& 120 		& 100 \\
\hline 
$2^3\P_0(3916)$	& DD		& 0.0 (see text) 		& 0 \\
		& $gg$		   	& 42. 		& 99.5 \\
		& $\psi'(2^3\S_1) \gamma$ & 0.14	& 0.3\\
		& $\psi_1(1^3\D_1) \gamma$ & 0.05 	& 0.1\\
		& Total			& 42 		& 100 \\
\hline 
$2^1\P_1(3956)$	& DD$^*$		& 78.4		& 97.9 \\
		& $ggg$		   	& 1.29 		& 1.6 \\
		& $gg\gamma$		& 0.13 		& 0.2 \\
		& $\eta_c'(2^1\S_0) \gamma$ & 0.22	& 0.3 \\
		& $\eta_c(1^1\S_0) \gamma$ & 0.08 	& 0.1 \\
		& Total			& 80 		& 100 \\
\hline \hline
\end{tabular}
\end{center}

\end{table}


\begin{thebibliography}{99}

\bibitem{Cho03}
S.K.Choi and S.L.Olsen  (Belle Collaboration),
[hep-ex/0309032].
%``Observation of a new narrow charmonium state in exclusive B+- 
% $\to$ K+- pi+ pi- J/psi decays,''
%%CITATION = HEP-EX 0309032;%%

\bibitem{Bau03}
G.Bauer (CDF Collaboration), Presentation at the 
{\it 2nd International 
Workshop on Heavy Quarkonium} (20-22 Sept. 2003, Fermilab).

\bibitem{Aub03}
B.Aubert {\it et al.}  (BABAR Collaboration),
Phys. Rev. Lett. 90, 242001 (2003)
[hep-ex/0304021].
%``Observation of a narrow meson decaying to D/s+ pi0 at a mass of  
%2.32-GeV/c**2,''
%%CITATION = HEP-EX 0304021;%%

\bibitem{Bes03}
D.Besson {\it et al.}  (CLEO Collaboration),
Phys. Rev. D68, 032002 (2003)
[hep-ex/0305100].
%``Observation of a narrow resonance of mass 2.46-GeV/c**2 
%decaying to  D/s*+ pi0 and confirmation of the D/sJ*(2317) state,''
%%CITATION = HEP-EX 0305100;%%

\bibitem{Kro03}
P.Krokovny {\it et al.} (Belle Collaboration)
hep-ex/0308019,
submitted to Phys. Rev. Lett.
% Title: Observation of the DsJ(2317) and DsJ(2457) in B decays
% Authors: P. Krokovny et al (Belle Collaboration)
% Comments: 6 pages, 3 figures, submitted to Phys. Rev. Lett., 
% Supercedes results reported in hep-ex/0307041
% SLAC-comments: Supersedes results reported in hep-es/0307041.

\bibitem{e705}
L.~Antoniazzi {\it et al.}  [E705 Collaboration],
Phys. Rev. D50, 4258 (1994).
%``Search for hidden charm resonance states decaying into J / Psi or Psi-prime
%plus pions,''
%%CITATION = PHRVA,D50,4258;%%

\bibitem{Abbrev}
In this paper an overbar is implicit 
if not explicitly present 
in two-meson open-charm final states. 
Thus ``DD" represents the state D${\bar {\rm D}}$, 
and ``DD$^*$" represents (D${\bar {\rm D}}^* + h.c.$). 

\bibitem{Chi98}
G. Chiladze, A.F. Falk, and A.A. Petrov,
Phys. Rev. D58, 034013 (1998).

\bibitem{Yua97}
F. Yuan, C-F. Qiao, and K-T Chao, 
Phys. Rev. D56, 329 (1997).

\bibitem{Ko97}
P.~w.~Ko, J.~Lee and H.~S.~Song,
Phys. Lett. B395, 107 (1997)
[hep-ph/9701235].
%``Color-octet mechanism in the inclusive D-wave charmonium 
%productions in  B decays,''
%%CITATION = HEP-PH 9701235;%%

\bibitem{Eic02}
E.J.Eichten, K.Lane and C.Quigg,
Phys. Rev. Lett. 89, 162002 (2002)
[hep-ph/0206018].
%``B-Meson Gateways to Missing Charmonium Levels"
%%CITATION = HEP-PH 0206018;%%

\bibitem{Tor03}
N.A.T\"ornqvist
[hep-ph/0308277].
%``Comment on the narrow charmonium state of Belle at 3871.8 MeV 
%as a deuson"
%%CITATION = HEP-PH 0308277;%%
                                                                                
\bibitem{Clo03b}
F.E.Close and P.R.Page,
Phys. Lett. B574, 210 (2003) 
[hep-ph/0309253].
%``The D*0 D0bar threshold resonance,''
%%CITATION = HEP-PH 0309253;%%

\bibitem{Vol03b}
M.B.Voloshin
[hep-ph/0309307].
%``Interference and binding effects in decays of possible molecular 
% component of X(3872)"
%%CITATION = HEP-PH 0309307;%%

\bibitem{Pak03}
S.Pakvasa and M.Suzuki,
[hep-ph/0309294].
%``On the hidden charm state at 3872 MeV''
%%CITATION = HEP-PH 0309294;%%

\bibitem{Won03}
C.-Y.Wong
[hep-ph/0311088].
%``Molecular States of Heavy-Quark Mesons"
%%CITATION = HEP-PH 0311088;%%

\bibitem{Bra03}
E.Braaten and M.Kusunoki,
[hep-ph/0311147].
%``Low-energy Universality and the New Charmonium Resonance at 3870 MeV"
%%CITATION = HEP-PH 0311147;%%

\bibitem{Swa03}
E.S.Swanson
[hep-ph/0311229]. (added in proof)
%``Short Range Structure in the X(3872)"
%%CITATION = HEP-PH 0311229;%%

\bibitem{Clo03a}
F.E.Close and S.Godfrey,
Phys. Lett. B574, 210 (2003) 
[hep-ph/0305285].
%``Charmonium hybrid production in exclusive B meson decays''
%%CITATION = HEP-PH 0305285;%%

% Spectroscopy: quark models

\bibitem{Eic76}
E.Eichten, K.Gottfried, T.Kinoshita, K.D.Lane and T.M.Yan,
Phys. Rev. Lett. 36, 500 (1976).
%``The Interplay Of Confinement And Decay In The Spectrum Of Charmonium,''
%%CITATION = PRLTA,36,500;%%

\bibitem{Eic78}
E.Eichten, K.Gottfried, T.Kinoshita, K.D.Lane and T.M.Yan,
Phys. Rev. D17, 3090 (1978); 
{\it err.} 
Phys. Rev. D21, 313 (1980).
%``Charmonium: The Model''
%%CITATION = PHRVA,D17,3090;%%

\bibitem{Eic80}
E.Eichten, K.Gottfried, T.Kinoshita, K.D.Lane and T.M.Yan,
Phys. Rev. D21, 203 (1980).
%``Charmonium: Comparison With Experiment,''
%%CITATION = PHRVA,D21,203;%%
     
\bibitem{God85}
S.Godfrey and N.Isgur,
Phys. Rev. D32, 189 (1985).
%``Mesons In A Relativized Quark Model With Chromodynamics,''
%%CITATION = PHRVA,D32,189;%%

\bibitem{Eic81}
E.Eichten and F.Feinberg,
Phys. Rev. D23, 2724 (1981).
%``Spin Dependent Forces In QCD,''
%%CITATION = PHRVA,D23,2724;%%

\bibitem{Ful91}
L.P.Fulcher,
Phys. Rev. D44, 2079 (1991).
%``Perturbative QCD, A Universal QCD Scale, Long Range Spin Orbit Potential, 
% And The Properties Of Heavy Quarkonia,''
%%CITATION = PHRVA,D44,2079;%%

\bibitem{Gup86}
S.N.Gupta, S.F.Radford and W.W.Repko, 
Phys. Rev. D34, 201 (1986).
%%CITATION = PHRVA,D34,201;%%

\bibitem{Ebe03}
D.Ebert, R.N.Faustov and V.O.Galkin,
Phys. Rev. D67, 014027 (2003)
[hep-ph/0210381].
%``Properties of heavy quarkonia and B/c mesons in the relativistic 
% quark model,''
%%CITATION = HEP-PH 0210381;%%

\bibitem{Zen94}
J.Zeng, J.W.Van~Orden and W.Roberts,
Phys. Rev. D52, 5229 (1995)
[hep-ph/9412269].
%``Heavy mesons in a relativistic model,''
%%CITATION = HEP-PH 9412269;%%

% Spectroscopy: LGT

\bibitem{Bal01}
G.S.Bali,
Phys. Rept. 343, 1 (2001)
[hep-ph/0001312]. 
%%CITATION = HEP-PH 0001312;%%

\bibitem{Bal03}
G.S.Bali,
[hep-lat/0308015]. 
%%CITATION = HEP-LAT 0308015;%%

\bibitem{Manke}
T.Manke, personal communication.

\bibitem{Che01b}
P.Chen, X.Liao and T.Manke,
Nucl. Phys. Proc. Suppl. 94, 342 (2001) 
[hep-lat/0010069].
%``Relativistic Quarkonia from anisotropic lattices"
%%CITATION = HEP-LAT 0010069;%%

\bibitem{Lia02}
X.Liao and T.Manke
[hep-lat/0210030].
%``Excited charmonium spectrum from anisotropic lattices"
%%CITATION = HEP-LAT 0210030;%%

\bibitem{Che01a}
P.Chen,
Phys. Rev. D64, 034509 (2001)
[hep-lat/0006019].
%``Excited charmonium spectrum from anisotropic lattices"
%%CITATION = HEP-LAT 0006019;%%

\bibitem{Aok02}
S.Aoki {\it et al.} (CP-PACS Collaboration),
Nucl. Phys. Proc. Suppl. 106, 364 (2002)
[hep-lat/0110129].
%``Charmonium spectrum from quenched QCD on anisotropic lattices"
%%CITATION = HEP-LAT 0110129;%%

\bibitem{Oka02}
S.Okamoto {\it et al.} (CP-PACS Collaboration),
Phys. Rev. D65, 094508 (2002)
[hep-lat/0112020].
%``Charmonium Spectrum from Quenched Anisotropic Lattice QCD"
%%CITATION = HEP-LAT 0112020;%%

% E1 transitions

\bibitem{God03}
S.Godfrey,
Phys. Lett. B568, 254 (2003) [hep-ph/0305122].
% Testing the Nature of the D_{sJ}^*(2317)^+ and D_{sJ}(2463)^+ States 
% Using Radiative Transitions
% Journal-ref: Phys.Lett. B568 (2003) 254-260
% DOI: 10.1016/j.physletb.2003.06.049
%%CITATION = HEP-PH 0305122;%%
 
\bibitem{BAr03a}
W.A.Bardeen, E.J.Eichten and C.T.Hill,
Phys. Rev. D68, 054024 (2003)
[hep-ph/0305049].
%``Chiral multiplets of heavy-light mesons,''
%%CITATION = HEP-PH 0305049;%%

\bibitem{Col03}
P.Colangelo and F.DeFazio,
Phys. Lett. B570, 180 (2003)
[hep-ph/0305140].
%``Understanding D/sJ(2317),''
%%CITATION = HEP-PH 0305140;%%

\bibitem{Kwo88a}
W.Kwong and J.L.Rosner,
Phys. Rev. D38, 279 (1988).
%``D Wave Quarkonium Levels Of The Upsilon Family,''
%%CITATION = PHRVA,D38,279;%%

\bibitem{Sie37}
A.J.Siegert,
Phys. Rev. 52, 787 (1937).
%``Note On The Interaction Between Nuclei And Electromagnetic Radiation,''
%%CITATION = PHRVA,52,787;%%

\bibitem{McC83}
R.McClary and N.Byers,
Phys. Rev. D28, 1692 (1983).
%``Relativistic Effects In Heavy Quarkonium Spectroscopy,''
%%CITATION = PHRVA,D28,1692;%%

\bibitem{Mox83}
P.Moxhay and J.L.Rosner,
Phys. Rev. D28, 1132 (1983).
%``Relativistic Corrections In Quarkonium,''
%%CITATION = PHRVA,D28,1132;%%

% PDG 

\bibitem{PDG02}
Particle Data Group, 
K.Hagiwara {\it et al.}, 
Phys. Rev. D66, 010001 (2002).
%%CITATION = PHRVA,D66,010001;%%

\bibitem{Skw03}
T.Skwarnicki, Presentation to the 
Lepton-Photon Conference, Fermilab August 2003.

% 3P0 decay model: Zweig allowed decays

\bibitem{Bar03b}
T.Barnes, 
to appear in Proceedings of HADRON'03
[hep-ph/0311102].
%``Strong Decays: Past, Present and Future"
%%CITATION = HEP-PH 0311102;%%

\bibitem{Ack96}
E.S.Ackleh, T.Barnes and E.S.Swanson,
Phys. Rev. D54, 6811 (1996) 
[hep-ph/9604355].
%%CITATION = HEP-PH 9604355;%%

\bibitem{Blu96}
A discussion of uncertainties in these decay models appears in
H.G. Blundell and S. Godfrey, 
Phys. Rev. D53, 3700 (1996)
[hep-ph/9508264].
%%CITATION = HEP-PH 9508264;%%

% Annihilation decays

\bibitem{App75}
T.Appelquist and H.D.Politzer, Phys. Rev. Lett. 34, 43 (1975);
%%CITATION = PRLTA,34,43;%%

\bibitem{DeR75}
A.DeRujula and S.L.Glashow, Phys. Rev. Lett. 34, 46 (1975).
%%CITATION = PRLTA,34,46;%%

\bibitem{Cha75}
M.Chanowitz, Phys. Rev. D12, 918 (1975).
%%CITATION = PHRVA,D12,918;%%

\bibitem{Bar76a}
R.Barbieri, R.Gatto and R. K\"ogerler, 
Phys. Lett. B60, 183 (1976).
%%CITATION = PHLTA,B60,183;%%

\bibitem{Bar76b}
R.Barbieri, R.Gatto and E.Remiddi, 
Phys. Lett. B61, 465 (1976).
%%CITATION = PHLTA,B61,465;%%

\bibitem{Nov78}
V.A.Novikov, L.B.Okun, M.A.Shifman, A.I.Vainshtein, M.B.Voloshin, 
and V.I.Zakharov, Phys. Rept. C41, 1 (1978).
%%CITATION = PRPLC,41,1;%%

\bibitem{Bar79}
R.Barbieri, G.Curci, E.d'Emilio and E.Remiddi, 
Nucl. Phys. B154, 535 (1979).
%%CITATION = NUPHA,B154,535;%%

\bibitem{Kwo88b} 
W.Kwong, P.B.Mackenzie, R.Rosenfeld, and J.L.Rosner,
Phys. Rev. D37, 3210 (1988).
%%CITATION = PHRVA,D37,3210;%%

\bibitem{Ber91}
L.Bergstr\"om and P.Ernstr\"om, 
Phys. Lett. B267, 111 (1991).
%%CITATION = PHLTA,B267,111;%%

\bibitem{Ack92a}
E.S.Ackleh and T.Barnes,
Phys. Rev. D45, 232 (1992). 
% TWO PHOTON WIDTHS OF SINGLET POSITRONIUM AND QUARKONIUM 
% WITH ARBITRARY TOTAL ANGULAR MOMENTUM.
% accuracy of contact formulas
%%CITATION = PHRVA,D45,232;%%

\bibitem{Ack92b}
E.S.Ackleh, T.Barnes and F.E.Close,
Phys. Rev. D46, 2257 (1992).
% TWO PHOTON HELICITY SELECTION RULES AND WIDTHS FOR POSITRONIUM AND 
% QUARKONIUM STATES WITH ARBITRARY ANGULAR MOMENTA.
% contact formulas
% ccbar annihilation transitions
%%CITATION = PHRVA,D46,2257;%%

\bibitem{Yan80}
T.M.Yan, Phys. Rev. D22, 1652 (1980).
%%CITATION = PHRVA,D22,1652;%%

\bibitem{Kua81}
Y.P.Kuang and T.M.Yan, 
Phys. Rev. D24, 2874 (1981).
%%CITATION = PHRVA,D24,2874;%%

\bibitem{Kua88}
Y.P.Kuang, S.F.Tuan and T.M.Yan, 
Phys. Rev. D37, 1210 (1988).
%%CITATION = PHRVA,D37,1210;%%

\bibitem{Kua90}
Y.P.Kuang and T.M.Yan,
Phys. Rev. D41, 155 (1990).
% ``Hadronic Transitions Of D Wave Quarkonium And Psi (3770) 
% $\to$ J / Psi Pi Pi,''
%%CITATION = PHRVA,D41,155;%%

\bibitem{Vol80} 
M.B.Voloshin and V.I.Zakharov, 
Phys. Rev. Lett. 45, 688 (1980).
%%CITATION = PRLTA,45,688;%%

\bibitem{Vol86} 
M.B.Voloshin, 
Sov. J. Nucl. Phys. 43, 1011 (1986).
%``Hadronic Transitions From Upsilon (3s) To 1 P Wave Singlet Bottomonium
%Level,''
%%CITATION = SJNCA,43,1011;%%

\bibitem{Iof80} 
B.L.Ioffe and M.A.Shifman, 
Phys. Lett. 95B, 99 (1980).
%%CITATION = PHLTA,B95,99;%%

\bibitem{Nov81} 
V.A.Novikov and M.A.Shifman, 
Z. Phys. C8, 43 (1981).
%``Comment On The Psi-Prime $\to$ J / Psi Pi Pi Decay,''
%%CITATION = ZEPYA,C8,43;%%

\bibitem{Vol03a} 
M.B.Voloshin 
Phys. Lett. B562, 68 (2003)
[hep-ph/0302261].
%%CITATION = HEP-PH 0302261;%%

\bibitem{Bel87}
G.Belanger and P.Moxhay, 
Phys. Lett. B199, 575 (1987). 
%``Three Gluon Annihilation Of D Wave Quarkonium,''
%%CITATION = PHLTA,B199,575;%%

\bibitem{Mox88}
P.Moxhay, 
Phys. Rev. D37, 2557 (1988).
%%CITATION = PHRVA,D37,2557;%%

\bibitem{Eic94}
E.J.Eichten and C.Quigg,
%``Mesons with beauty and charm: Spectroscopy,''
Phys. Rev. D49, 5845 (1994)
[hep-ph/9402210].
%%CITATION = HEP-PH 9402210;%%

\bibitem{Bar03a}
T.Barnes and N.I.Kochelev,
[nucl-th/0306026].
%``Closed-flavor pi + J/psi and pi + Upsilon Cross Sections 
%at Low Energies from Dipion Decays"
%%CITATION = NUCL-TH 0306026;%%

\bibitem{Bai03}
J.Z.Bai {\it et al.}  (BES Collaboration)
[hep-ex/0307028].
%``Evidence of psi(3770) non-(D anti-D) decay to J/psi pi+ pi-,''
%%CITATION = HEP-EX 0307028;%%

\bibitem{Voloshin}
M.B. Voloshin,  comment at the 
{\it 2nd International 
Workshop on Heavy Quarkonium} (20-22 Sept. 2003, Fermilab).

% early charm molec refs

\bibitem{Vol76}
M.B.Voloshin and L.B.Okun,
JETP Lett. 23, 333 (1976).
%``Hadron Molecules And Charmonium Atom,''
%%CITATION = JTPLA,23,333;%%

\bibitem{DeR76}
A.De~Rujula, H.Georgi and S.L.Glashow,
Phys. Rev. Lett. 38, 317 (1977).
%``Molecular Charmonium: A New Spectroscopy?,''
%%CITATION = PRLTA,38,317;%%

\bibitem{Rosner}
If the $X(3872)$ is a ${\rm DD}^*$ molecule it is quite possibly a 
member of an I-spin multiplet whose charged members may decay to 
$J/\psi\, \rho^\pm \to J/\psi\, \pi^\pm\pi^0$.  
We thank J.Rosner for suggesting this possibility.

% LGT hybrids

\bibitem{McN02}
C.McNeile {\it et al.} (UKQCD Collaboration),
Phys. Rev. D65, 094505 (2002) 
[hep-lat/0201006].
%``Hybrid meson decay from the lattice"
%%CITATION = HEP-LAT 0201006;%%

\end{thebibliography}
\end{document}